\documentclass[traditabstract]{aa} 
\usepackage{graphicx}
\usepackage{lscape}
\usepackage{txfonts}
\usepackage{natbib}
\usepackage{url}
\usepackage[T1]{fontenc}

\usepackage[ps2pdf,pdfpagemode=UseThumbs,colorlinks=true,bookmarks=true]{hyperref}
\hypersetup{citecolor={blue},linkcolor={blue},urlcolor={blue}}

\newcommand{\EWHa}{EW($H\alpha$)~}

\newcommand{\fudenlowSpx}{10$^{-18}$ erg~s$^{-1}$~cm$^{-2}$~spaxel$^{-1}$}

\newcommand{\degree}{\ensuremath{^\circ}}

\newcommand{\nii}{[\ion{N}{ii}]}

\newcommand{\oiii}{[\ion{O}{iii}]}
\newcommand{\sii}{[\ion{S}{ii}]}
\newcommand{\siii}{[\ion{S}{iii}]}
\newcommand{\Ariii}{[\ion{Ar}{iii}]}
\newcommand{\lam}{$\lambda$}

\newcommand{\ha}{H$\alpha$} 
\newcommand{\hb}{H$\beta$}

\DeclareRobustCommand{\ion}[2]{%
\relax\ifmmode
\ifx\testbx\f@series
{\mathbf{#1\,\mathsc{#2}}}\else
{\mathrm{#1\,\mathsc{#2}}}\fi
\else\textup{#1\,{\mdseries\textsc{#2}}}%
\fi}

\newcommand{\HII}{\ion{H}{ii}~}

\begin{document}
   \title{Census of \HII regions in NGC 6754 derived with MUSE: Constraints on the metal mixing scale.} 


   \author{
     S.\,F. S\'anchez\inst{\ref{unam}}
     \and
     L. Galbany\inst{\ref{mil},\ref{chile}}
     \and
     E. P\'erez \inst{\ref{iaa}}
     \and
     P.\, S\'anchez-Bl\'azquez\inst{\ref{uam}}
     \and
     J. Falc\'on-Barroso \inst{\ref{iac},\ref{lalag}}     
     \and
     F.\,F. Rosales-Ortega\inst{\ref{inaoe}}
     \and
     L.\, S\'anchez-Menguiano\inst{\ref{iaa},\ref{ugr}}
     \and 
     R.\, Marino\inst{\ref{ucm}}
     \and
     H. Kuncarayakti\inst{\ref{mil},\ref{chile}}
     \and
     J. P. Anderson\inst{\ref{eso}}
     \and
     T. Kruehler\inst{\ref{eso}}
     \and
     M. Cano-D\'\i az\inst{\ref{unam}}
     \and
     J. K. Barrera-Ballesteros\inst{\ref{iac},\ref{lalag}}
     \and
     J. J. Gonz\'alez-Gonz\'alez\inst{\ref{unam}}
          }


   \institute{
        \label{unam}Instituto de Astronom\'\i a, Universidad Nacional Auton\'oma de M\'exico, A.P. 70-264, 04510, M\'exico, D.F.
     \and
     \label{mil}Millennium Institute of Astrophysics, Universidad de Chile, Casilla 36-D, Santiago, Chile
     \and
     \label{chile}Departamento de Astronom\'\i a, Universidad de Chile, Casilla 36-D, Santiago, Chile
        \and
        \label{iaa}Instituto de Astrof\'{\i}sica de Andaluc\'{\i}a (CSIC), Glorieta de la Astronom\'\i a s/n, Aptdo. 3004, E-18080 Granada, Spain
        \and
       \label{uam}Departamento de F\'isica Te\'orica, Universidad Aut\'onoma de Madrid, 28049 Madrid, Spain.
        \and
        \label{iac}Instituto de Astrof\'\i sica de Canarias (IAC), E-38205 La Laguna, Tenerife, Spain 
        \and
        \label{lalag}Depto. Astrof\'\i sica, Universidad de La Laguna (ULL), E-38206 La Laguna, Tenerife, Spain
        \and
       \label{inaoe}Instituto Nacional de Astrof{\'i}sica, {\'O}ptica y Electr{\'o}nica, Luis E. Erro 1, 72840 Tonantzintla, Puebla, M\'exico
       \and
       \label{ugr}Dpto. de F\'\i sica Te\'orica y del Cosmos, University of Granada, Facultad de Ciencias (Edificio Mecenas), E-18071 Granada, Spain
       \and
\label{ucm}CEI Campus Moncloa, UCM-UPM, Departamento de Astrof\'{i}sica y CC$.$ de la Atm\'{o}sfera, Facultad de CC$.$ F\'{i}sicas, Universidad Complutense de Madrid, Avda.\,Complutense s/n, E-28040 Madrid, Spain.
       \and
\label{eso}European Southern Observatory, Alonso de Cordova 3107 Casilla 19001 - Vitacura -Santiago, Chile.
              }

   \date{Received ----- ; accepted ---- }

 
\abstract{ 

We present a study of the \HII regions in the galaxy NGC 6754 
from a two pointing mosaic comprising 197,637 individual spectra, 
using Integral Field Spectrocopy (IFS) recently acquired with the MUSE instrument during its
Science Verification program. The data cover the entire galaxy out to $\sim$2 effective radii ($r_e$), 
sampling its morphological structures with unprecedented spatial resolution for { a wide-field IFU.}
A complete census of the \HII regions
limited by the atmospheric seeing conditions was derived, comprising
396 individual ionized sources. This is one of the largest and
most complete catalogue of \HII regions with spectroscopic information
in a single galaxy. We use this catalogue to derive the radial
abundance gradient in this SBb galaxy, finding a negative gradient with
a slope consistent with the characteristic value for disk
galaxies recently reported. The large number of \HII regions
allow us to estimate the typical mixing scale-length ($r_{mix}\sim$0.4
$r_e$), which sets strong constraints on the proposed mechanisms for
metal mixing in disk galaxies { , like radial movements associated with bars and spiral arms, when comparing with simulations}. 
{ We found evidence for an azimuthal variation of the oxygen abundance,
that may be related with the radial migration.} These results illustrate the unique
capabilities of MUSE for the study of the enrichment mechanisms in
Local Universe galaxies.}


\maketitle


\section{Introduction}
\label{intro}

Nebular emission lines have been historically the main tool at our
disposal for direct measurement of the gas-phase abundance at discrete
spatial positions in low-redshift galaxies
\citep[e.g.][]{allo79}. They trace the young, massive star component
in galaxies, illuminating and ionizing cubic kiloparsec-sized volumes
of interstellar medium. Metals play a fundamental role in cooling
mechanisms in the intergalactic and interstellar medium, and in processes of 
star-formation, stellar physics, and planet formation. 

Previous spectroscopic studies have unveiled some aspects of the
complex processes at play between the chemical abundances of galaxies
and their physical properties. These studies have been successful in
determining important relationships, scaling laws and systematic
patterns
\citep[e.g.][]{leque79,diaz89,zaritsky94,Garnett:2002p339,tremonti04,moustakas06}. However,
these results are limited by statistics, either in the number of
observed \ion{H}{ii} regions or in the coverage of these regions
across the galaxy surface.

The advent of multi-object spectrometers and IFS instruments with large fields of view (FoV) now offers
the opportunity to undertake a new generation of emission-line
surveys, based on samples of hundreds of \ion{H}{ii} regions and full
two-dimensional (2D) coverage of the disks of nearby spiral galaxies
\citep[e.g.][]{rosales-ortega10}.
One of the most interesting results recently derived using IFS data
is that the oxygen abundance gradient seems to present a common slope
$\sim$$-$0.1 dex/$r_e$ for non-interacting galaxies \citep{sanchez12b,sanchez14}.

This result agrees with models based on the standard inside-out
scenario of disk formation, which predict a relatively quick self
enrichment with oxygen and an almost universal negative
metallicity gradient once it is normalized to the galaxy optical
size \citep{bois99,bois00}.  From the seminal works of \cite{lace85},
\cite{guest82} and \cite{clay87}, most numerical models of chemical evolution
 explain the existence of the radial gradient of abundances by
the combined effects of a star-formation rate and an infall of gas,
both varying with galactocentric radius  \citep[e.g.,][]{molla99}.

{ Although there is a large number of studies focused on the analysis
of the abundance gradient in galaxies, in contrast, little is known about
the possible presence of azimuthal asymmetries in this distribution. Deviations
from the radial abundance gradient are well known features in the Milky Way, based on the
study of Cepheids and open clusters \cite[e.g.,][]{chiap01,lepi11}. However, the situation in other spiral galaxies is less clear, and suffers from poor statistics, either for the low number of \HII regions sampled
per galaxy or for the large errors of the abundance estimation. Recently, using
wide-field IFS \cite{rosales11} showed that the radial metallicity gradient of NGC 628 varies
slighly for different quadrants, although the differences are comparable to the uncertainties
introduced by the adopted estimators of the oxygen abundances. More recently, \cite{li13}
found marginal evidence for the existence of moderate deviations from chemical abundance homogeneity
in the intestellar medium of M101, using a combination of strong-line abundance indicators
and direct estimations based on the detection of the \oiii$\lambda$4363 auroral line.}

Despite the advances of recent IFS-surveys in our understanding of
the evolution of the chemical enrichment processes in galaxies, they
present some limitations.  The most important one is the lack of
the spatial resolution required to propertly resolve individual
small-scale morphological structures, in particular individual \HII
regions. The IFS surveys with the best physical resolution, such as PINGS
\citep{rosales-ortega10} or CALIFA \citep{sanchez12a}, have $\sim$5
times worst spatial resolution than the typical ground-based imaging
surveys. This results in a bias in the detection of \HII regions, that
are aggregated based on their spatial vicinity (decreasing their number
by a factor three or more), and their spectra are polluted by diffuse
gas emission \citep{mast14}.

\begin{figure*}
\centering
\includegraphics[width=18.2cm]{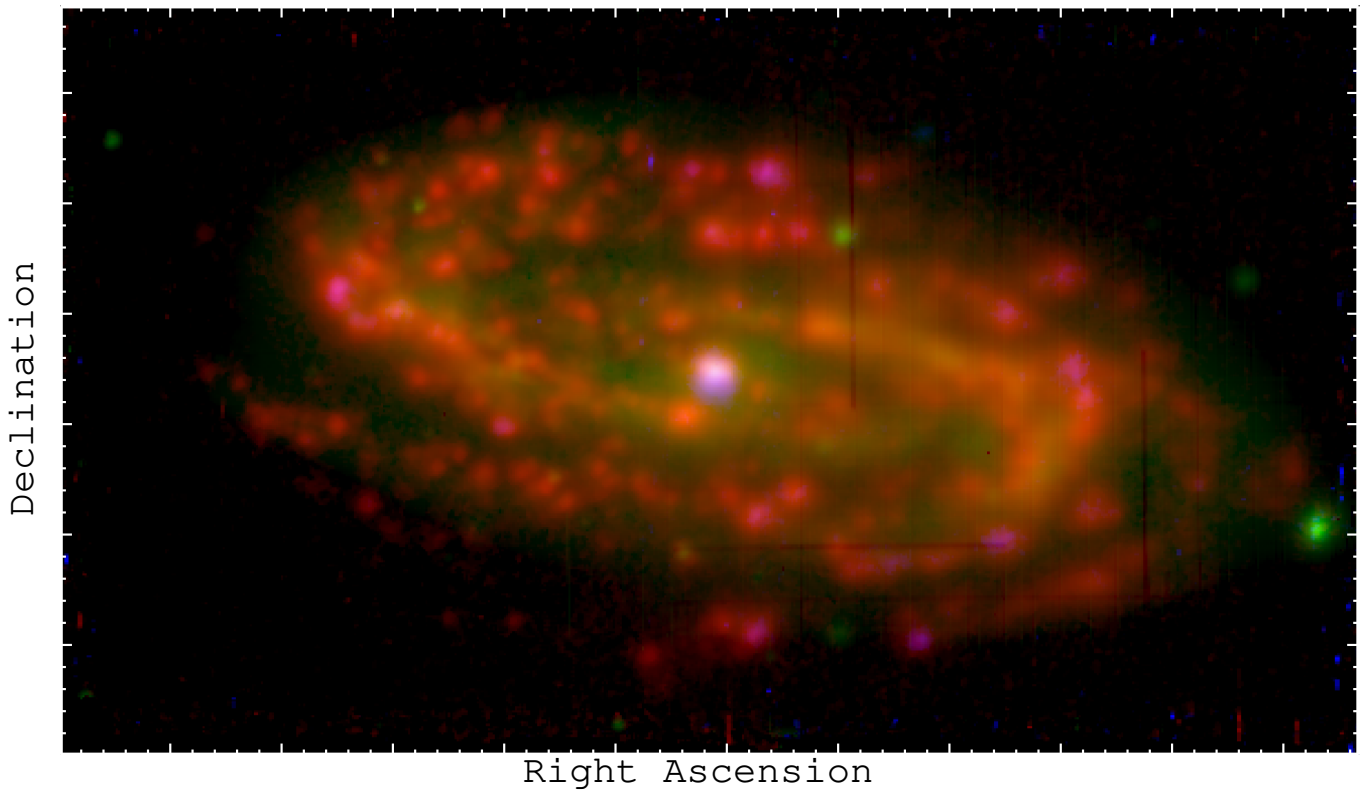}
\caption{\label{fig:color} RGB color image of NGC 6754 created using the line intensity maps of [\ion{O}{iii}]$\lambda$5007 (blue), $V$-band (green) and H$\alpha$ (red) extracted from the datacubes. Each large tickmark corresponds to 10$\arcsec$ (or 400 pc at the redshift of the galaxy). The green point-like sources are field stars. Each red structure corresponds to a single \HII region.}
\end{figure*}

In principle, the abundance scatter { and azimuthal asymmetries} of  \HII regions
around the average radial gradient can be used to constrain the spatial-scale of radial mixing 
\citep[e.g.][]{scalo04,dimat13}. In the absence of radial mixing the
only observed scatter around the abundance gradient should be produced
by the errors in the individual measurements. Regardless of its origin \citep[e.g.,][]{atha92}, 
any radial mixing increases the scatter by moving regions of a
certain abundance from a certain galactocentric distance to
a different one. Therefore, the dispersion around the average slope is a
constraint to the maximum radial mixing scale.

So far, for the reasons outlined above, the current IFS surveys lacked 
the required resolution to address this important key issue in the
chemical evolution of galaxies. The Multi Unit Spectroscopic Explorer \citep[MUSE][]{MUSE} 
has changed dramatically the perspective for these studies. This instrument is a
unique tool for the spectroscopic analysis of resolved structures in
galaxies, particularly in the local universe. The combination of a
large { field-of-view (FoV)} ($\sim$60$\arcsec$$\times$60$\arcsec$), unprecedented spatial
sampling (0.2$\arcsec$/spaxel) { for a wide-field IFU}, which limits the spatial resolution to
the atmospheric seeing, the spectral resolution and large
wavelength coverage, and the large aperture of the VLT telescope, makes
MUSE a well suited instrument to address these { problems}.
{ Of course, there are other IFUs with similar or even larger FoVs, 
like PPAK \citep{kelz06}, VIMOS \citep{vimos} or VIRUS-P \citep{virusP}, and also other
IFUs operating in the optical range have similar or better spatial sampling, like GMOS \citep{gmos}
or OASIS\footnote{\url{http://cral.univ-lyon1.fr/labo/oasis/present/}}. However, MUSE is the first one that combines at the
same time the large FoV and the image-like spatial sampling.}

In this work we study the oxygen abundance gradient of the spiral
galaxy NGC 6754 using the data recently observed by MUSE as part of the Science
Verification programs (SV). NGC 6754 is a barred Sb galaxy
mildly inclined ($i\sim$60$\degree$). Its brightness (B$\sim$13 mag),
projected size (r$_{25}\sim$ 1$\arcmin$) and redshift ($z=$0.0108),
similar to the footprint of the CALIFA galaxies
\citep[e.g.][]{walcher14}, makes it suitable to perform a census of the
\HII regions using MUSE, to further understand the 
abundance distribution in this galaxy.

\section{Data acquisition and reduction}\label{data}

NGC 6754 was observed on June 28th and 30th 2014 in the context of
Program 60.A-9329 (PI: Galbany) of the MUSE SV
run. The observations were divided into two pointings covering the
east and west parts of the galaxy, respectively. The final cube for
each pointing is the result of 3 exposures of 900 seconds, where the
second and third exposure were slightly shifted (2 arcsec NE and SW,
respectively) and rotated 90$\degree$ from the first exposure, in order to
provide a uniform coverage of the field and to limit systematic errors in
the reduction.

The reduction of the raw data was performed with {\tt Reflex} \citep{Reflex} using
version 0.18.2 of the MUSE pipeline \citep{MUSEPIPE}, including the
standard procedures of bias subtraction, flat fielding, wavelength
calibration, flux calibration, and the final cube reconstruction by
the spatial arrangement of the individual slits of the image slicers. 

The final dataset comprises two cubes of $\sim$100k individual
spectra, each covering a FoV slightly larger than $\sim$1$^{\square}$. 
Each spectrum covers the wavelength range 4800-9300 \AA, 
with a typical spectral resolution between 1800 and 3600 (from
blue to red). The cubes are aligned east-west, with an overlapping area
of $\sim$16$\arcsec$, where the galaxy center has been sampled
twice.  The final mosaiced datacube comprises almost 200k individual
spectra, covering the entire galaxy up to 2 effective radii, with a
FoV of $\sim$2$\arcmin$$\times$1$\arcmin$. For practical reasons, we
analysed each cube separately and later combined the different
data products.

Figure \ref{fig:color} illustrates the power of the combined large FoV and
high spatial resolution of MUSE, and the quality of the data. It shows
a true color image created using a combination of a $V$-band image,
and two continuum subtracted narrow-band images of { 30\AA} width, 
centred in [\ion{O}{iii}]$\lambda$5007 and H$\alpha${, at the
redshift of the galaxy. The three maps were synthetized from each datacube,
and combined to create a single image for each band. For each of the narrow-band images the continuum
was estimated from the average of two additional narrow-band images of
similar width (i.e., 30\AA) extracted at a wavelength redshifted and blueshifted 100\AA\ from
the nominal wavelength of the considered emission line at the redshift of the object. Finally,
the $V$-band image was synthetized by convolving the spectra at each spaxel by the
nominal response curve of the Johnson $V$-band filter. Those images were used to illustrate the spatial
resolution and image quality of the data. We should note here than the $V$-band image does not include
only continuum emission, since it is contaminated by both H$\beta$ and [\ion{O}{iii}]$\lambda$5007, and
the H$\alpha$ intensity map is contaminated by the adjacent [\ion{N}{ii}] doublet. However, they
clearly trace the continuum and emission line distribution across the galaxy. } The spatial distribution of the
individual \HII/ionized regions can be easily recognized, tracing the star forming regions 
along the spiral arms of the galaxy. { Different seeing conditions for the for the East ($\sim$0.8$\arcsec$) and for the West ($\sim$1.8$\arcsec$) pointings are also clear.}

\section{Analysis}\label{ana}

The main goals of this study are to characterize the abundance
gradient in NGC 6754 and to estimate the dispersion of the abundances of
the individual \HII regions around the average gradient. In this section we
describe briefly how we select the \HII regions, extract and
analyze their individual spectra, derive the corresponding oxygen
abundance, and analyze their radial gradient. More details
on the procedure are described in \cite{sanchez12b}, and references therein.

\subsection{Detection of the ionized regions}\label{HII_detect}

The segregation of \ion{H}{ii} regions and the extraction of the
corresponding spectra is performed using a semi-automatic procedure
named {\sc  HIIexplorer}\footnote{\url{http://www.caha.es/sanchez/HII_explorer/}}.
The details of this program are given in \cite{sanchez12b}, and a
detailed description of the overall detection process,
\cite{sanchez14}. {  {\sc  HIIexplorer} requires as input} a map of emission line
intensities or equivalent widths, { a minimum threshold above which
  the peak intensity of the \HII region is detected, and three
  different convergence criteria: (i) the maximum fractional
  difference between the peak intensity and the adjacent ones to be
  agregated to a particular region, (ii) the minimum absolute
  intensity for a pixel to be agregated, and (iii) the maximum distance between
  the  pixel considered and the peak intensity} .  In this
particular case we use the map of the equivalent width of H$\alpha$,
EW(${H\alpha}$), derived from the narrow-band image described
above. { The use of the EW guarantees that the analysis is more
  homogeneous between the two pointings, since this parameter is less
  affected by possible spectrophotometric differences and it is less
  sensitive to seeing variations. The fact that the equivalent width
  may be contaminated or not by the adjacent [\ion{N}{ii}] doublet is not
  relevant, since this map is used only to detect the emission line
  regions, and not in any further analysis along the
  article. Therefore, a possible contamination by [NII] may affect
  only the contrast, and only marginally, but
  not the detectability of the regions, since the average
  contamination by this line is about a 30\% of the total flux.} The output of {\sc
  HIIexplorer} is a segmentation map and the integrated spectrum for
each of the \HII regions detected.

\begin{figure*}
\centering
\includegraphics[angle=270,width=18.2cm]{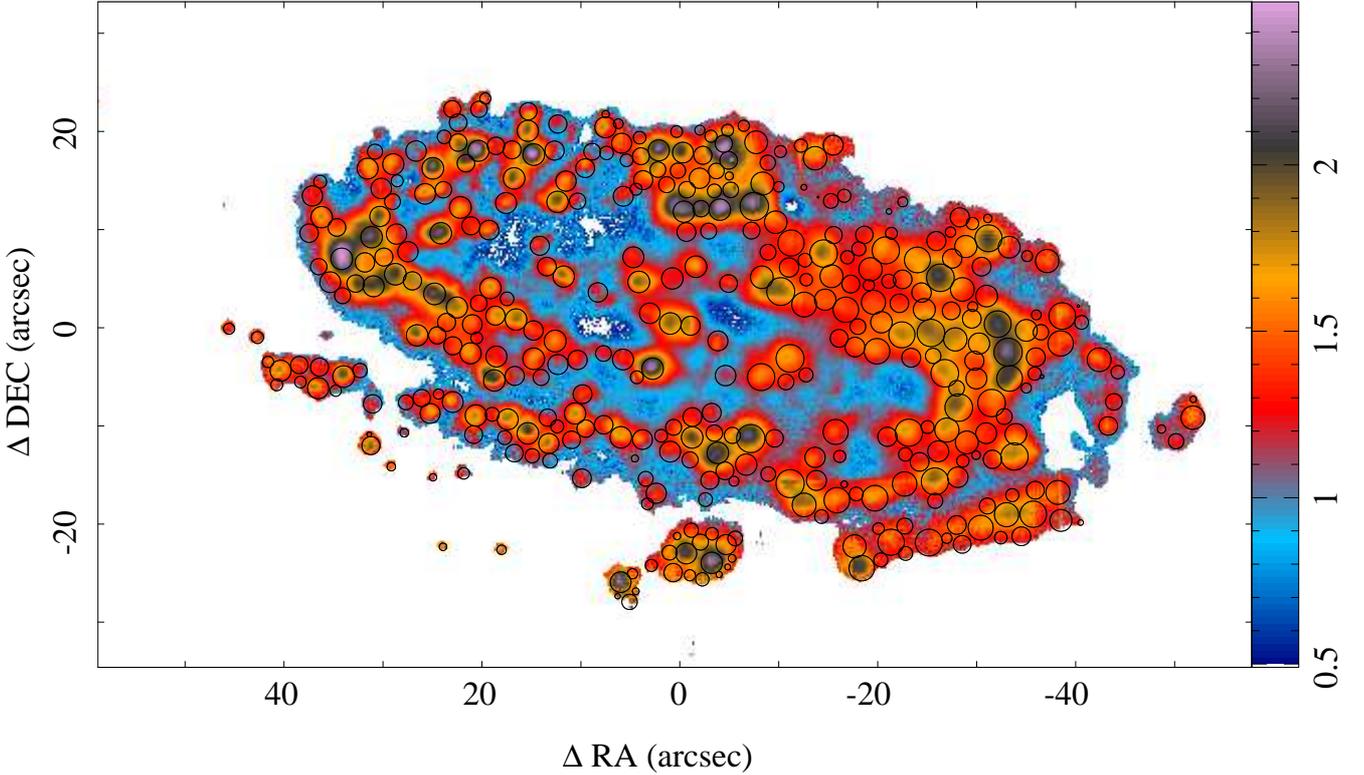}
\caption{\label{fig:EW} Color coded map of the equivalent width of H$\alpha$ in logarithmic scale. The areas with  H$\alpha$ density flux below 1.5 \fudenlowSpx ($\sim$3$\sigma$ detection limit) have been masked. The circles represent the detected \HII regions, width the radius proportional to the extraction aperture.}
\end{figure*}

We processed individually the two datacubes, { fixing the input parameters to
the optimal ones for the west-pointing, that was observed under worst
seeing conditions. We selected a threshold in the peak  \EWHa = 20\AA, and minimum 
\EWHa = 8\AA, a minimum fractional peak of 1\%, and a maximum distance of 2\arcsec. 
Therefore, the convergence criteria restrict the detection of regions with at least
 \EWHa = 8\AA\ in every spaxel and a maximum diameter of 4\arcsec. The selection
of these parameters is based on our previous studies with other IFU data and different
tests to optimize the results: (i) the minimum absolute \EWHa is selected to guarantee
that all the pixels agregated to a particular region are above the boundary between
retired and star-forming regions proposed by \cite{cid-fernandes10} and discussed
in \cite{sanchez14}, even if a very conservative error of 25\% is assumed for this parameter,
and/or taking into account the contamination by [\ion{N}{ii}]; (ii) 
the threshold in the peak \EWHa is selected to be more than
twice the minimum, to guarantee that the region is actually clumpy/peaky, and
not a diffuse ionized region; (iii) the maximum distance is fixed to 
the estimated size of an \HII region, that could have a diameter as large as $\sim$1 kpc \citep[e.g., NGC\,5471, ][]{oey03,rgb2011}, thus, $\sim$4\arcsec at the redshift of the object. Using these parameters
we detect a similar number of \HII regions in each pointing: 207 in the east pointing and
220 in the west one.}

\begin{figure*}
\centering
\includegraphics[width=7cm,angle=270]{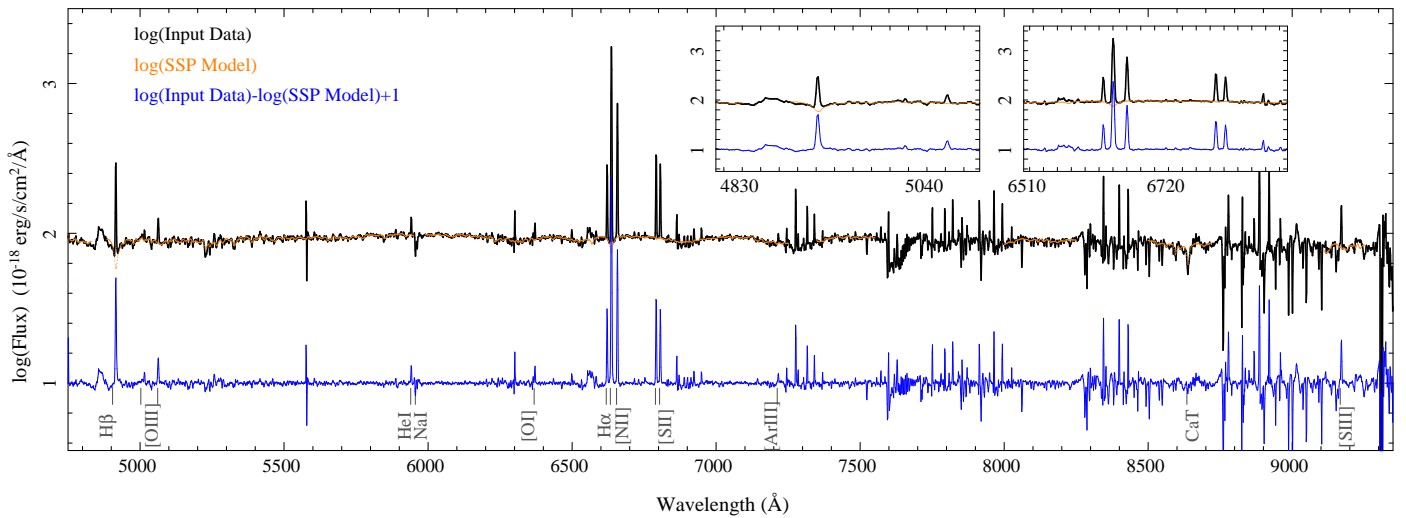}
\caption{\label{fig:spec}  Detail of the spectrum of a typical \HII region extracted from the galaxy. The black line shows the input spectrum in logarithmic scale, together with the best fitted stellar model, in orange. The difference between the logarithms of the input spectrum and the model, shifted by 1, is shown as a blue line. The intensities of the emission lines are so high that the spectra are plotted in logarithmic scale to show them together with the detail of the underlying stellar population. { The most prominent spectral features discussed along the article are marked. The spectral regions masked during the fitting of the underlying stellar population are not shown in the orange-solid line. The two boxes show an expanded view around H$\beta$ and H$\alpha$ respectively.}}
\end{figure*}

The final catalogue was cleaned for double detections { in the
  overlapping area} by removing those \HII regions with coordinates
that differ less than 3$\arcsec$. A total of 396 individual clumpy
ionized regions are detected, a factor of 5-10 times larger than the
number found with lower spatial resolution IFU data
\cite[e.g.][]{sanchez13}, as predicted by the simulations presented by
\cite{mast14}. { Figure \ref{fig:EW} shows the  EW(${H\alpha}$) map illustrating the result of this
  procedure.  The
  location and relative size of the  \HII regions detected are indicated with
  black  circles. We note here that the {\sc HIIexplorer}
  provides with a segmentation map, not with circular apertures. The
  current representation is therefore illustrative of the size of the \HII regions, but
  does not show the actual detailed shape of the associated segmented regions for which the spectra are
  extracted.}

\subsection{Measurement of the emission line intensities}\label{HII_elm}

In this analysis we follow the procedures described in
\cite{sanchez14}, using the fitting package {\tt
  FIT3D}\footnote{\url{http://www.caha.es/sanchez/FIT3D/}},(\cite{sanchez06b,sanchez11}).  We perform a Monte-Carlo fitting using { two
  different single stellar population (SSP) libraries}.  { In order
  to compare with previous results and provide with useful information
  of the underlying stellar population, we first use a library that
  comprises 156 templates to model and remove the underlying stellar
  population. This library comprises 39 stellar ages, from 1 Myr to 13
  Gyr, and 4 metallicities ($Z/Z_{\odot}=$ 0.2, 0.4, 1, and 1.5), and
  it is described in detail in \cite{cid-fernandes13}. These templates
  were extracted from { a combination of the synthetic stellar spectra
  from the GRANADA library \cite{martins05} and } the SSP library provided by the MILES project
  \citep{miles,vazdekis10,falc11}. Therefore they are restricted to a
  wavelength range lower than 7000\AA. Hence, they cannot be used to
  remove the underlying stellar population in the full spectral range
  covered by MUSE. For doing so, we used a more restricted library
  extracted from the MIUSCAT models \citep{miuscat}. It comprises 8
  stellar ages, from 65~Myr to 17.7~Gyr, and 3 metallicities
  ($Z/Z_{\odot}=$ 0.4, 1, and 1.5). Our previous experience indicates
  that to decouple the underlying stellar population from the emission lines
  a restricted library like this one is enough \citep[e.g.][]{sanchez14b}. 
  We do not find large differences between the residual spectra for the
  wavelength range in common, and therefore we finally adopted the results from the second library
  for the analysis of the emission lines.}

 Dust attenuation and stellar
kinematics were taken into account as part of the fitting process.
{ The stellar kinematics was derived as a first step, fitting the underlying
stellar population with a sub-set of the full stellar library changing
the systemic velocity and the velocity dispersion at random within the
range of allowed values. Then a first model for the stellar population
is derived. This model is used to obtain the dust attenuation, 
allowed to change randomly within a pre-defined range. The extinction law
by \cite{cardelli89} was assumed, with a specific dust attenuation of
$R_{\rm V}=3.1$. For each iteration over the dust attenuation values a new
model of the underlying stellar population is derived. The best combination
of the stellar velocity, velocity dispersion and dust attenuation is recovered
based on the lowest reduced $\chi^2$ provided. Finally, these parameters are fixed and
the full stellar library is used to recover the underlying stellar population.}
{ Due to the  Monte-Carlo fitting adopted, the inaccuracies in the derivation of the underlying
stellar population are propagated to the error budget in the emission line fitting, and therefore, into
the errors estimated for the emission line fluxes.}

Individual emission line fluxes were measured in the {\it
  stellar-population subtracted} spectra by fitting each of them with
a single Gaussian function. For this particular dataset we extracted
the flux intensity of the following emission lines: \ha, \hb,
\oiii\ \lam4959, \oiii\ \lam5007, \nii\ \lam6548, \nii\ \lam6583, {
  \sii\ \lam6717, \sii\ \lam6731, \Ariii\ \lam7135 and
  \siii\ \lam9069}. We may notice here that many other relevant
emission lines are detected within the wavelength range covered by the
spectra.  Figure \ref{fig:spec} shows a detail of the spectrum of a
typical \HII region within the sample, together with the best fitted
stellar population model and the resulting emission-line spectrum. In
addition to the emission lines measured, it is possible to identify
weaker lines like He\,{\footnotesize I} \lam5876 and
[O\,{\footnotesize I}] \lam6300. The two components of the
Na\,{\footnotesize I} absorption doublet at $\sim$5892\AA\ are clearly
resolved, with a contribution due to attenuation that cannot be
reproduced by the stellar model, which was actully masked during the
fitting process.{ The prominent CaT$\lambda$8542\AA\ stellar
  absorptions in the near-infrared is also clearly detected. In
  addition to the emission line fluxes, we derive the emission line
  equivalent widths. For doing so, we divide the integrated flux of
  the line by the median flux density of the best fitted SSP-model in
  a window of 100\AA\ centred in the wavelength of the emission line.
  Therefore, these equivalent widths are not contaminated by the
  contribution of any adjacent emission lines.}

\begin{figure}
\centering
\includegraphics[width=8.5cm,angle=270]{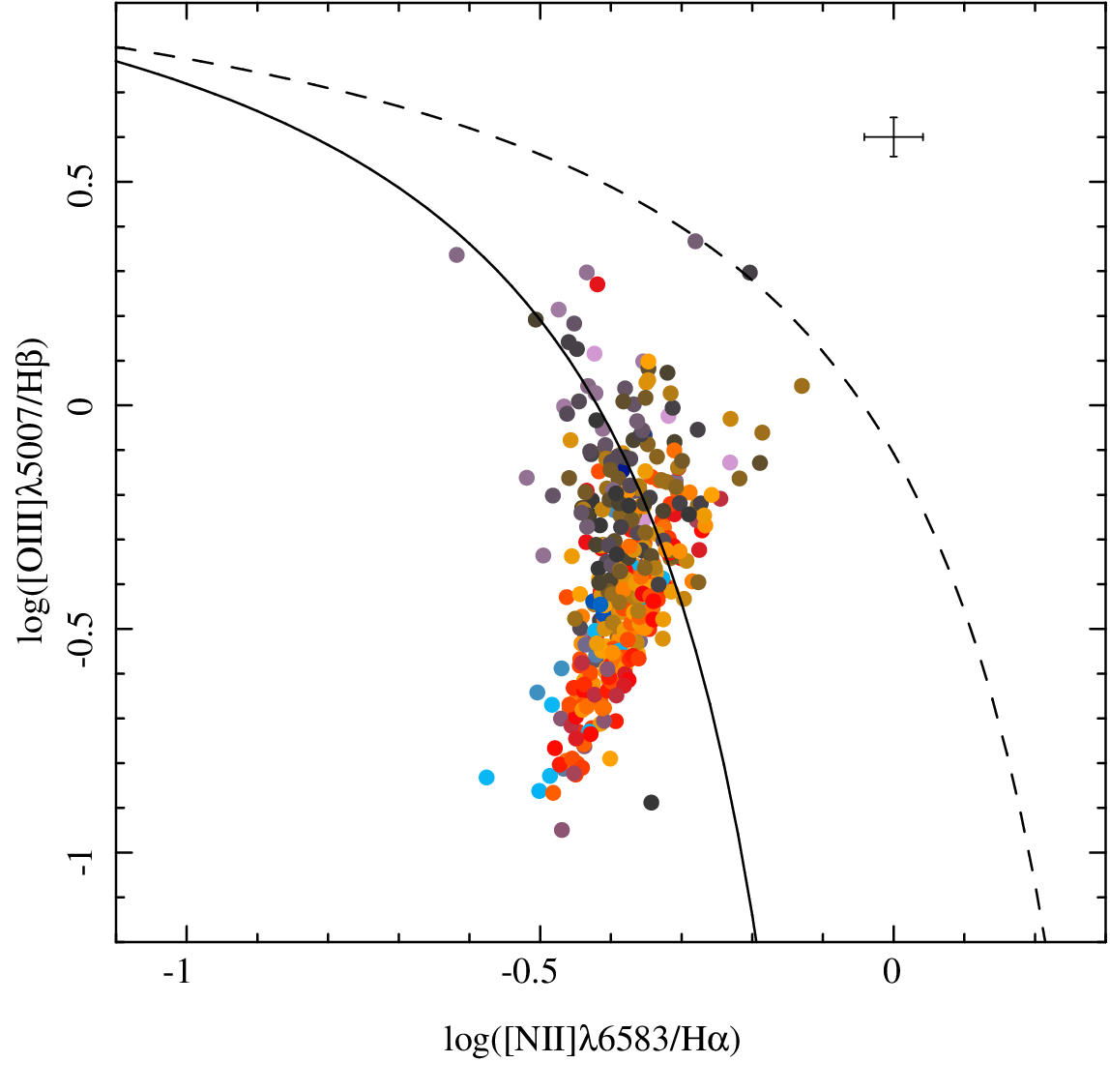}
\caption{\label{fig:diag}  [\ion{O}{iii}]~$\lambda$5007/H$\beta$ vs. [\ion{N}{ii}]~$\lambda$6583/H$\alpha$ diagnostic diagram for the 396 \HII ionized regions detected in NGC 6754, color coded by the deprojected galactocentric distance (where bluer colors correspond to the central regions, and reddish-to-grey colors correspond to the outer regions). Solid and dashed lines represent, respectively, the \cite{kauffmann03} and \cite{kewley01} demarcation curves. They are usually invoked to distinguish between classical star-forming objects (below the { solid} line), and AGN powered sources (above the dashed line). Regions between both lines are considered intermediate ones. The average error of the line ratios is represented by the error bar in the upper-right corner. }
\end{figure}

{ Two  artifacts in the spectra are  identified in
  Fig. \ref{fig:spec}, in particular in the residual emission line spectrum). 
They look like two bumps at $\sim$4750\AA\ and
  $\sim$6550\AA, just bluewards of H$\beta$ and H$\alpha$. Although
  they have been masked during the fitting process we prefer to show
  them in the plot, since they seem to be present in all the MUSE spectra we
  have analyzed. We are not sure if they are a product of the
  reduction process or a feature in the raw data. In any case, they do
  not affect the measurements of the emission lines.  Two additional
  spectral features that are not well reproduced by the SSP models
  correspond to the \ion{Na}{i}$\lambda$5890,5896 absorption feature,
  at $\sim$5950\AA\ at the redshift of the object, and the
  CaT$\lambda$8542\AA, at $\sim$8630\AA\ at the redshift of the
  object. The \ion{Na}{i} mismatch is a well known feature since this
  absorption line has two physical origins: (i) the absorption due to
  the presence of this element in the atmosphere of the stars, which
  is included in the SSP-templates, and (ii) the absorption due to the
  presence of this element in the inter-stellar medium, which produces
  an absorption proportional to the gas content (and dust
  attenuation). The CaT$\lambda$8542\AA\ is often problematic in the
  current SSP-templates, that could be related to variations in the
  IMF, the abundance of Ca, or a non correct understanding of this
  absorption feature (although it is not the case in the particular
  example shown in Fig.\ref{fig:spec}, for which we show the
  prediction from the fitted SSP-model). Due to these well known
  mismatches, both spectral regions were masked during the fitting
  process. Residuals from non perfect sky-subtraction of the strong
  OH-lines in the near-infrared are visible in the redder wavelength
  ranges, and a clear defect associated with a telluric absoption is
  shown at $\lambda$$\sim$7600\AA. None of the emission lines considered
  in this study are affected by this later effect.}


%
\subsection{Selection of \HII regions}\label{HII_select}

Classical \ion{H}{ii} regions are gas clouds ionized by short-lived
hot OB stars, associated with ongoing star-formation. They are
frequently selected on the basis of demarcation lines defined in the
so-called diagnostic diagrams \citep[e.g.,][]{baldwin81,veilleux87},
which compare different line ratios. Figure \ref{fig:diag} shows the
classical diagram using [\ion{O}{iii}]/H$\beta$ vs. [\ion{N}{ii}]/H$\alpha$
\citep[][BPT diagram hereafter]{baldwin81}, for the sample of \HII
regions described above. This diagnostic diagram is frequently used
since it uses very strong emission lines (e.g., Fig.\ref{fig:spec}),
and it is less affected by dust attenuation and imperfections in the
spectrophotometric calibration. The classical demarcation lines described by
\cite{kauffmann03} and \cite{kewley01} have been included. 

In this study we follow \cite{sanchez13} to select \HII
regions as clumpy ionized regions with EW(H$\alpha$)$>$6\AA\ and
located below the \cite{kewley01} demarcation curve.  This
selection guarantees the exclusion of ionized regions possibly  
dominated by shocks (that are not clumpy in general), regions whose
ionization is dominated by post-AGB stars, and AGN-dominated regions
\citep[e.g.][]{cid-fernandes10}. Following these criteria all the
ionized regions selected by {\tt HIIexplorer} have been classified as
\HII regions.

The distribution of \HII regions across the BPT diagram follows the
characteristic pattern in Sa/Sb early-type spirals \cite{sanchez14b}.
They are mostly located at the bottom-right end of the classical
location of \HII regions in this diagram, with a tail towards the
so-called intermediate region between the two demarcation lines
described above. \cite{kennicutt89} first recognized that \ion{H}{ii}
regions in the center of galaxies are spectroscopically different from
those in the disk in their stronger low-ionization forbidden emission,
that place them in the so-called intermediate region. More recently
\cite{sanchez14} found a similar behaviour for the \HII regions in
early-type disk galaxies, like NGC 6754. The location of the \HII
regions change across the BPT diagram with the galactocentric distance
(Fig. \ref{fig:diag}), as a consequence of the change of the
ionization conditions and in particular the radial gradient in the
oxygen abundance \cite[e.g., ][]{evans85,dopita86}.

{ 

\subsection{Physical conditions in the \HII regions}\label{phy}

The location in the BPT diagram for a classical \HII region ionized by
young stars of a certain age is defined well by three main parameters:
(a) the ionization parameter or fraction of Lyman continuum photons
with respect to total amount of gas, (b) the electron density of the
gas, and (c) the metallicity or chemical abundance of the ionized gas.
We derive here the first two of them. In addition, we estimate the
dust attenuation to correct the emission line fluxes 
when required.

The dust attenuation, A$_{\rm V}$, was derived for each \HII region
based on the H$\alpha$/H$\beta$ Balmer line ratio. The extinction law
by \cite{cardelli89} was assumed, with a specific dust attenuation of
$R_{\rm V}=3.1$, and the theoretical value for the unobscured line
ratio for case B recombination of H$\alpha$/H$\beta=2.86$, for
$T_e$=10,000\,K and $n_e$=100\,cm$^{-3}$ \citep{osterbrock89}.  For
this study we have assumed that the intrinsic H$\alpha$/H$\beta$ line
ratio does not vary significantly, although it is known that it
presents a dependence with the electron density and the temperature
\citep[e.g.][]{osterbrock89}. Once derived the dust attenuation,
all the considered emission line fluxes were corrected adopting the
same extinction law, when needed.

The electron density, $n_e$, was derived from the 
line ratio of the [\ion{S}{ii}] doublet \citep[e.g.,][]{osterbrock89}, by solving
the equation,

\begin{eqnarray}\label{eq2}
 \frac{I({\rm [\ion{S}{ii}]\lambda6717})}{I({\rm [\ion{S}{ii}]\lambda6731})} &=& 1.49 \frac{1+3.77{\rm x}}{1+12.8{\rm x}}
\end{eqnarray}
where $x$ is the density parameter, defined as $x=10^{-4} n_e
t^{-1/2}$ and $t$ is the electron temperature in units of $10^4$~K
\citep{McCa85}. For this calculation we assumed a typical electron
temperature of $T=10^4$K, which is an average value that corresponds to the
expected conditions in \HII regions \citep{osterbrock89}. This equation
reflects that the [\ion{S}{ii}] doublet ratio is sensitive to
changes in the electron density only for a limited range of
values. For high and low values, it becomes asymptotic, and the 
value derived  has to be treated with care and should not be used for
quantitative statements. However, the value will still be valid to understand
the possible dependences of the abundance gradient with this parameter.

For the ionization parameter, $u$, we adopted the
[\ion{S}{iii}]$\lambda$9069,9532/[\ion{S}{ii}]$\lambda$6717,6731
calibrator described by \cite{kewley02}. Since
[\ion{S}{iii}]$\lambda$9532 is not covered by our wavelength range, we
adopted a theoretical ratio of
[\ion{S}{iii}]$\lambda$9532/[\ion{S}{iii}]$\lambda$9069 = 2.5
\citep{pepe96}, fixed by atomic physics. Both
emission lines were corrected for the dust attenuation prior to deriving
the ionization parameter.

}

\begin{figure*}
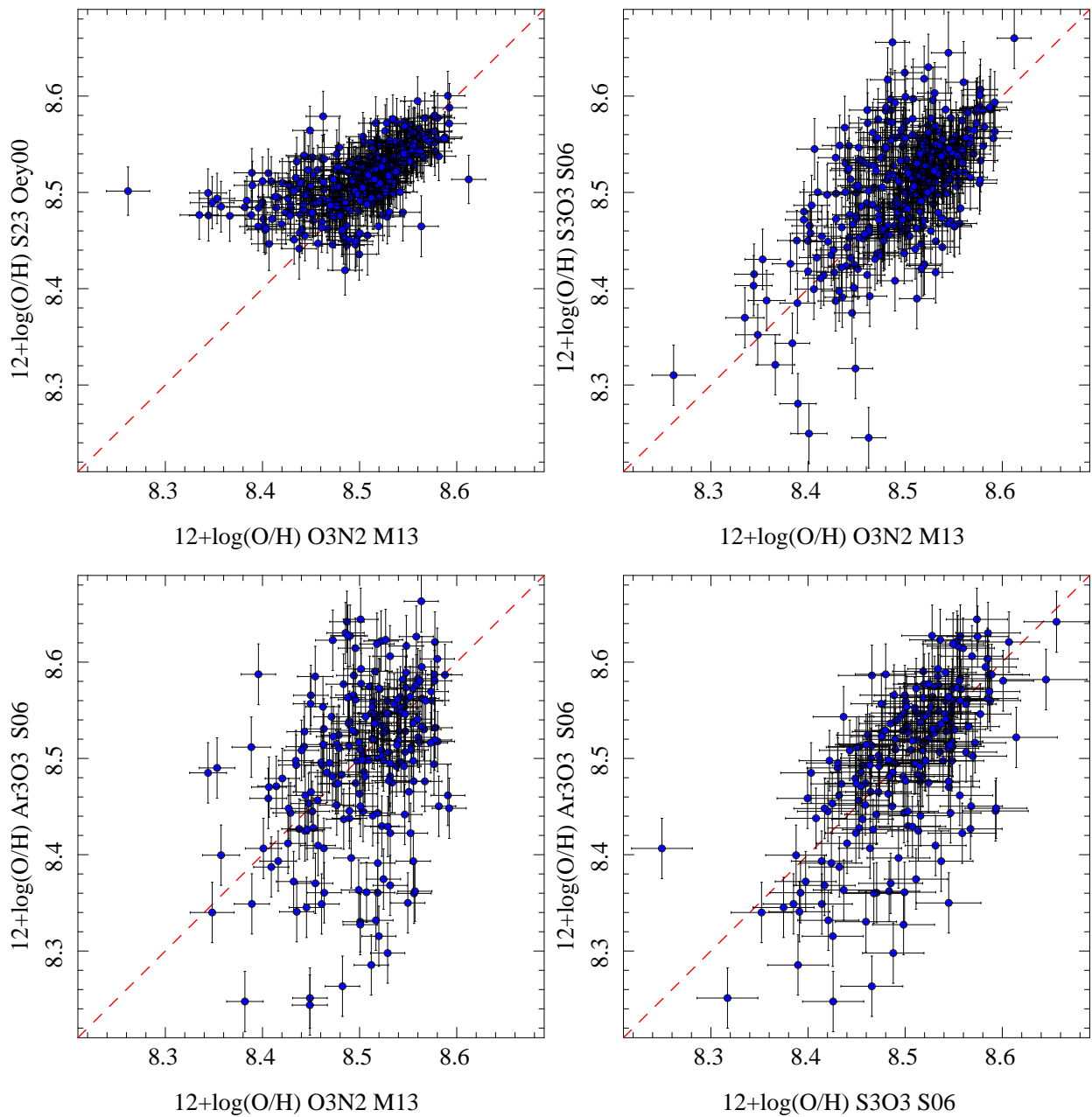

\centering
\includegraphics[angle=270,width=8.2cm]{figs/OH_O3N2_S23.ps}\includegraphics[angle=270,width=8.2cm]{figs/OH_O3N2_S3O3.ps}
\includegraphics[angle=270,width=8.2cm]{figs/OH_O3N2_Ar3O3.ps}\includegraphics[angle=270,width=8.2cm]{figs/OH_S3O3_Ar303.ps}
\caption{\label{fig:comp} Comparison among the oxygen abundances derived using the different indicators described in the text: (i) S23 vs. O3N2 (top lefthand panel); (ii) S3O3 vs. O3N2 (top righthand panel); (iii) ArO3 vs. O3N2 (bottom lefthand panel); and (iv) ArO3 vs. S3O3 (bottom righthand panel). Each blue solid circle corresponds to an individual \HII region in the sample. The error bars represent half of the estimated error considering the propagation of the emission line errors and the uncertainty in the calibrator. The dashed-line corresponds in each panel to the one-to-one relation.}
\end{figure*}

\subsection{Oxygen abundance of \HII regions}\label{HII_OH}

{ Accurate abundance measurements for the ionized gas in galaxies
  require the determination of the electron temperature (T$_e$), 
usually obtained from the ratio of auroral to
  nebular line intensities \citep[e.g.][]{osterbrock89}.  It is well
  known that this procedure is difficult to carry out for metal-rich
  galaxies, since as the metallicity increases the electron
  temperature decreases (as the cooling is via metal lines), and the
  auroral lines eventually become too faint to measure. Therefore,
  calibrators based on strong emission lines are used. 

Strong-line indicators have the obvious advantage of using emission
lines with higher signal-to-noise, detected in mostly all \HII
regions, and with large dynamical ranges. However, they have the
disadvantange that the  line ratios considered do not trace only the
oxygen abundance, but also depend on other properties of the ionized nebulae, like
the electron density, and geometrical factors, and/or the shape of the
ionizing radiation (normally parametrized by the ionization
parameter, $q$, or $u=q/c$ in its dimensionless form). It is well known that some of these parameters
are correlated, like the trend between oxygen abundance and
the ionization parameter, uncovered by the seminal studies by \cite{evans85,dopita86},
and recently revisited by \cite{sanchez14b}.

There are  two main schools in the derivation of oxygen abundance using
 strong-line indicators. One  uses empirical calibrators based
 on the comparison of different of strong emission line ratios,
 with the corresponding abundance derived for a set of \HII regions 
 for which T$_e$ is known. The line ratios in these methods are:

\begin{eqnarray}
 {\rm R23} &=& \frac{I([\ion{O}{iii}]~\lambda\lambda5007,4959)+I([\ion{O}{iii}]~\lambda3727)}{I({\rm H}\beta)} \\
 {\rm O3N2} &=& \frac{I([\ion{O}{iii}]~\lambda5007)/I({\rm H}\beta)}{I([\ion{N}{ii}]~\lambda6584)/I({\rm H}\alpha)} \\
 {\rm N2O2} &=&  \frac{I([\ion{N}{ii}]~\lambda6584)}{I([\ion{O}{ii}]~\lambda3727)} \\
   {\rm N2} &=&  \frac{I([\ion{N}{ii}]~\lambda6584)}{I(H\alpha)}.
\end{eqnarray}

 This school adopts in most cases emission line ratios that present
 little or no dependence with the dust attenuation (i.e., not far in
 wavelength), that uses the strongest available emission lines, 
 and calibrators that present either a monotonic or even a linear dependence with
 the abundance. For example, the O3N2 and the N2 indicators
 \citep{allo79,pettini04,stas06,marino13}. In some cases they use a
 combination of all the available emission lines, like the {\it
   counterpart}-method described by \cite{P12}, or more complex
 combinations of non-linear equations to derive the abundances
 \citep[e.g.][]{pilyugin10,epm14}. Since they adopted empirical correlations, the
 intrinsic dependences with other parameters, like $u$, are subsummed
 in the calibrator by construction.

A second school prefers to use photoionization models to derive the
dependence of the abundance and ionization strength with the different
line ratios \citep[e.g.][]{dopita00,kewley01}. A certain model for the
ionizing stellar population is assumed, taking into account a certain
burst of star formation, an initial mass function, a certain
metallicity and, in some cases, the age of the cluster. Under
certain conditions for the ionized nebulae (e.g., geometry, electron
density), it is possible to derive trends and correlations between the
abundance and the  line ratios considered. For this school, the
prefered line ratios are those that depend only of one of the
parameters, or for which the correction on the other is well
understood.  The procedure/prescriptions described by \cite{kewley02}
or \cite{angel12} on how to derive the abundances is a good example of
this approach. 

The main difference between the two schools is that for the first one the
abundances derived  are systematically lower. \cite{dopita14} studied
a scenario in which the introduction of a $\kappa$ distribution for
the electron temperatures in the nebula allows to reconcile both
estimates of the oxygen abundance, based on the early studies by
\cite{binn09}. However, most of the followers of the first school
still consider that the T$_e$ or direct method is more representative
of the physical conditions in the nebulae, and requires fewer
assumptions or dependences on still not well understood physical
properties (like the amount of ionizing photons of young stars,
that differs among the different stellar evolution models).  Another
criticism is that in many cases the calibrations based on photoionization
models assume tight or fixed correlations between the abundance of
different elements (like the N/O ratio), that may affect their
derived values \citep[e.g.][]{epm14}.

We adopted the indicator based on the O3N2 ratio described before, in
order to compare with previous results on the same field.  This line
ratio involves the stronger emission lines in the 
wavelength range, as clearly appreciated in Fig. \ref{fig:spec}, and
therefore minimize the errors due to the inaccuracies in the
measurement of the involved line ratios. By construction, it presents a
weak dependence on dust attenuation, already noticed by previous
authors \citep[e.g.][]{kewley02}.  We adopted the recently updated
calibration by \citet{marino13}, that uses the largest sample of \HII
regions with abundances derived using the direct, T$_e$-based,
method (M13 hereafter). This calibration corrects the one proposed by
\citet{pettini04}, that (due to the lack of \HII regions in the upper abundance range) 
combined direct measurements for the lower
abundance range and values derived from photoionization models for the
more metal rich ones.  As demonstrated by \citet{marino13}, it
produces abundance values very similar to the ones estimated based on
the \cite{P12} method, with an accuracy better than $\pm$0.08 dex.

The wavelength range covered by MUSE at the redshift of the galaxy
does not include the [\ion{O}{ii}]$\lambda$3727 emission line.
Therefore, all indicators that include it, like R23, N2O2, or the combination
of any of them, can not be used. However, there are other less common
abundance indicators covered in this wavelength range, such as:

\begin{eqnarray}
 {\rm S23} &=& \frac{I([\ion{S}{iii}]~\lambda\lambda9069,9532)+I([\ion{S}{ii}]~\lambda6717,6731)}{I({\rm H}\beta)} \\
 {\rm S3O3} &=& \frac{I([\ion{S}{iii}]~\lambda\lambda9069)}{I([\ion{O}{iii}]~\lambda5007)}\\
 {\rm Ar3O3} &=& \frac{I([\ion{Ar}{iii}]~\lambda\lambda7135)}{I([\ion{O}{iii}]~\lambda5007)}.
\end{eqnarray}

The S23 indicator was first proposed by \cite{diaz00b} as an alternative
to the more widely used R23. Its main advantages are that the intensity
of the  lines are less affected by dust attenuation, that it presents a monotonic linear
dependence with the abundance for a wide range of metallicities, and
that it seems to be less dependent on the ionization parameter. This
later statement was questioned by \cite{kewley02} on the basis of
photoionization models. \cite{oey00} already noticed that this
indicator has a bi-valued behaivour with respect to the abundance,
similar to R23, and restricted the use of the calibrator proposed by
\cite{diaz00b} to subsolar metallicities ($Z<0.5 Z_{\cdot}$). This
corresponds to an oxygen abundance of 12+log(O/H)$<$8.3, lower than
the lowest abundances derived for the \HII regions discussed here
based on the M13 calibrator.

There is no published calibration of the dependence of S23 with the
oxygen abundance for the higher abundance branch. However, based on the
photoionization models presented by \cite{oey00}, it is possible to
derive an estimation of the abundance for that range:

\begin{eqnarray}
{\rm 12+log(O/H) }= 8.6 - 0.25 {\rm log} ({\rm S23}) 
\end{eqnarray}

The accuracy of this calibrator has to be tested extensively, but based
on the range of values covered by the described models we estimate to be
not better than 0.15 dex.

The S3O3 and Ar3O3 indicators were proposed by \cite{sta06} (S06 hereafter). She
derived a non-linear correlation for both indicators and the oxygen
abundance, that we adopt here. She estimated the accuracy of this
calibrator to be of the order of $\sim$0.09 dex. 

In contrast with the O3N2 abundance indicator, that is basically
independent of dust attenuation, all these indicators involve
ratios of emission lines widely separated in wavelength, and
therefore the line intensities must be corrected for dust attenuation
prior to derive the corresponding ratio and abundance. This introduces
a new degree of uncertainty that we avoid by adopting the O3N2 calibrator.

Figure \ref{fig:comp} shows the comparison among the different
estimators for the oxygen abundance discussed. We consider
only the emission lines and line ratios detected above a 5$\sigma$
detection limit.  Hence, each panel shows a
different number of \HII regions, ranging between 210 (for the
panels involving the ArO3 indicator), and 360 (for the panels involving
the O3N2, S23 and S3O3 indicators).  This is because the \Ariii\
emission line is fainter than any of the other (Fig.\ref{fig:spec}). 
There is very good agreement between the different estimators, 
despite the different ions involved in most of
the calibrators, and the inhomogenous derivation of the calibrators.  
The largest differences are found in the calibrator involving the
\Ariii\ emission line, for the reason indicated before: $\sigma(X_{\rm O3N2}-X_{\rm Ar3O3}) = 0.08$ and $\sigma(X_{\rm S3O3}-X_{\rm Ar3O3}) = 0.07$ dex, where $\sigma$ is
    the standard deviation of the difference between the two
    estimations of the abundance, and $X$ is the oxygen abundance, i.e., 12+log(O/H).  The smallest differences are found
    between the O3N2 and S23 indicators, with $\sigma(X_{\rm O3N2}-X_{\rm S23}) = 0.04$ dex, a value smaller than
      the expected accuracies of both calibrators. In summary, 
this comparison shows that our estimation of the oxygen abundance
does not depend strongly on the adopted indicator, and that in average
the accuracy of our estimation is of the order or better than $\sim$0.05 dex.
This systematic error has been included in the error budget of the abundances
derived for each individual \HII region.}

\subsection{Structural parameters of the galaxy}\label{morph}

We derive the mean position angle, ellipticity, and effective radius
of the disk, by a surface brightness and morphological analysis
performed on the MUSE data using a V-band image of NGC 6754
synthetized from the IFS cube.  { The procedure is extensively
  described in \cite{sanchez14}. In summary, an
  isophotal analysis is performed using the {\tt ellipse\_isophot\_seg.pl} tool
  included in the \textsc{HIIexplorer}
  package\footnote{\url{http://www.caha.es/sanchez/HII_explorer/}}.
  Unlike other tools, like {\tt ellipse} included in IRAF, this
  tool does not assume \emph{a priori} a certain parametric shape for
  the isophotal distributions. The following procedures were followed
  for the V-band image: (i) the peak intensity emission within a
  certain distance of a user defined center of the galaxy was
  derived. Then, any region around a peak emission above a certain
  percentage of the galaxy intensity peak is masked, which effectively
  masks the brightest foreground stars; (ii) once the peak intensity
  is derived, the image is segmented in consecutive levels following a
  logarithmic scale from this peak value; (iii) once the image is
  segmented in $n_{levels}$ isophotal regions, for each of them, a set
  of structural parameters was derived, including the mean flux
  intensity and the corresponding standard deviation, the semi-major
  and semi-minor axis lengths, the ellipticity, the position-angle,
  and the barycenter coordinates. The median values of the derived
  position angles and ellipticities along the different isophotes,
  once excluded those affected by the seeing in the very central
  regions, are adopted as the position angle and ellipticity of the
  galaxy. Their standard deviations are considered as an estimation of
  the error in the derivation of these parameters. We derive an
  ellipticity of $e=$0.88$\pm$0.07 and a position angle of
  ${\rm PA}=$77$\pm$9$\degree$.  Assuming an intrinsic ellipticity for
  the galaxy of $\sim$0.13 \citep{giov95,giov97}, the inclination is
  estimated to be $i=$64$\pm$6$\degree$. Finally, we fit the surface brightness
profile with a single exponential function to derive the disk scale-length, and
the correponding disk effective radius, as defined by \cite{sanchez14}. The
 effective radius derived at the distance of the galaxy was estimated as $r_e=$10.7$\pm$0.9 kpc }.

\section{Results}\label{res}

\subsection{Oxygen abundance gradient}\label{OH_grad}

\begin{figure}
\centering
\includegraphics[width=7cm,angle=270]{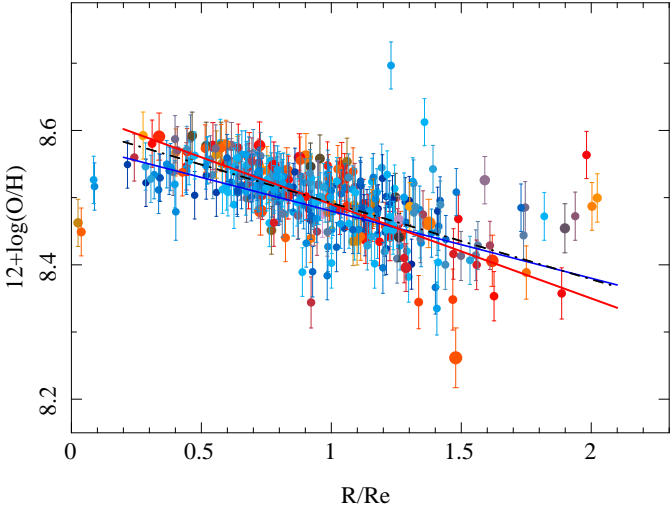}
\caption{\label{fig:grad} Radial distribution for the oxygen
  abundance derived using the O3N2 indicator for the 396 \HII regions
detected in NGC 6754, color coded by the equivalent width of H$\alpha$.
The blue colors and small symbols correspond to \HII regions with an emission EW(H$\alpha$) lower than 20\AA, while the reddish-to-grey colors and larger symbols correspond to regions with higher values. The error-bars illustrate the individual errors of the abundance propagated from the estimated errors of the emission lines, not including the systematic error of the abundance calibrator. The dashed-dotted black line shows the result of the best linear regression to all the points. The blue and red solid lines correspond to the results of the best linear regressions to the \HII regions of lower and higher values of EW(H$\alpha$).}
\end{figure}

We deproject the position of each \HII region using the morphological
parameters described in the previous section. Then, we derive the
galactocentric radial distribution of the oxygen abundance for NGC
6754, based on the abundances measured for each individual \ion{H}{ii}
region.  For the 396 \HII regions detected, figure \ref{fig:grad}
shows the abundance gradient derived out to $\sim$2 $r_e$ along the
galactocentric distance normalized to the effective radius.

{ The shape of the abundance gradient shown in this figure is
  totally consistent with the pattern found in many previous studies.
  The gradient shows an almost linear decrease between $\sim$0.3 and
  $\sim$1.7 effective radius, with a drop in the central regions and a
  flattenning and/or up-turn in the outer regions.  The linear regime has
  been interpreted as evidence of inside-out growth in spiral
  galaxies, with a metal enrichment dominated by local processes
  \citep[e.g.,][and references therein]{sanchez14}.  The drop in the
  inner region is found in a fraction of the spiral galaxies.  In some
  cases (e.g., NGC 628) it has been associated with a circumnuclear
  ring of star formation at the expected location of the inner
  Lindblad resonance radius, where the gas is indeed expected to
  accumulate, due to non-circular motions exerted by a bar or spiral
  arms \citep{sanchez11,rosales11}. The nature of the flattening in the
  outer regions, that has also been observed in other galaxies 
  \citep[e.g.,][]{mari11}, is still under debate. It could be an
  effect of the radial migration of stars that latter pollute the
  surrounding gas, or a consequence of a change in the
  star-formation efficiency. }

The dashed-dotted black line in Fig. \ref{fig:grad} shows
the error-weighted linear fit to this radial distribution of the
abundance.  Following \cite{sanchez14}, the analysis is restricted to
galactocentric distances 0.3$<r/r_e<$2.1.  We find a slope in the
abundance gradient of $\alpha$=$-$0.10$\pm$0.02 dex/$r_e$, which { is
  similar to } the common abundance slope reported by \cite{sanchez14}
of $\sim$$-$0.1 dex/$r_e$. { If we used the same calibrator than the
  one adopted in that study, i.e., \cite{pettini04}, instead of
  \cite{marino13}, the slope would be slightly larger,
  $\alpha$=$-$0.14$\pm$0.02 dex/$r_e$. In any case, both slopes are
  totally compatible with the common gradient, since they are both
  within 1$\sigma$ of the range of values described for this
  characteristic slope \citep{sanchez14}.}

The number of \HII regions detected for this galaxy is large enough to
explore whether the abundance gradient depends on other properties of
the nebular emission. The \HII regions in Fig. \ref{fig:grad} have
been color-coded according to the value of the
EW(${H\alpha}$). Adopting this scheme it is possible to distinguish
between the regions with stronger specific star-formation rate, that
trace mostly the spiral arms (Fig. \ref{fig:color}), and those
distributed more homogeneously across the entire disk. We
repeat the fitting procedure { splitting the sample in two, for \HII
  regions with equivalent width greater or smaller than 20\AA}. We
find that the \HII regions with lower equivalent widths present a {
  somewhat} shallower gradient ($\alpha$=$-$0.09$\pm$0.01) than those
with higher equivalent widths ($\alpha$=$-$0.12$\pm$0.02). The two
gradients are shown in Fig. \ref{fig:grad}. { However, the
  difference is rather small, and it may not be significant. In order
  to test it, we perform a Kolmogorov-Smirnov (KS) test to estimate how different the
  distributions of oxygen abundances in both cases are. We find that the
  probability that both distributions were not derived from the same
  sample is just a 7.7\%. Therefore, there is no significant difference
  between the abundance gradients for the regions with stronger or
  fainter specific star formation rates.

We explore possible differences in the abundance gradients based on
other properties of the ionized nebulae. First, we considered the
ionization parameter, splitting the sample in two subsamples with
log(u) greater or smaller than $-$3.6 (the median value for our
sample). The differences in the slopes were even smaller, being
$\alpha_{high log(u)}= -$0.10$\pm$0.02 dex and $\alpha_{low log(u)}=
-$0.08$\pm$0.03 dex. Then, we considered the electron density,
splitting the sample in regions with $n_e$ greater or smaller than 75
cm$^{-3}$ (the median value for our sample). In this case we find the
same slope for both subsamples ($\alpha_{high n_e}= -$0.10$\pm$0.03
dex).

Finally, we explore if the different spatial resolutions of the east
and west pointings have an impact in the abundance distributions and
gradients. We repeat the analysis restricting our sample to the
regions detected in both pointings separately before joining them into
a single catalog. We find very similar slopes for both subsamples:
$\alpha_{east}= -$0.10$\pm$0.02 dex and $\alpha_{west}=
-$0.12$\pm$0.03 dex. Even more, a KS-test indicates that the probability
that both distributions were not derived from the sample sample is 
0.02\%.

}

\subsection{Mixing scale-length}\label{OH_mix}

This sample of \HII regions is large enough to derive a 
estimation of the mixing scale-length. For doing so we compute the
dispersion of galactocentric distances with respect to the linear regression. 
The average of this relative distance
is zero (by construction), and the standard deviation is the typical mixing scale,
i.e., how far a certain \HII region has moved from its expected location
based on a pure inside-out chemical enrichment without radial mixing.
We find a mixing scale-length $r_{mix}=$0.43$r_e$, that corresponds to 4.6 kpc at the
redshift of this galaxy.  

{ We repeated the estimation for the different sub-samples of \HII
  regions discussed before, and found similar dispersions, covering a
  range of values of $r_{mix}=$0.37$-$0.53$r_e$. In particular, when
  taking into account the east and west pointing separately we
  derive a very similar radial mixing scale-length, slightly lower
  than the common one ($r_{mix,east/west}=$0.35$r_e$). This indicates
  that (i) the different spatial resolution does not affect the result,
  and (ii) there seems to be an azimuthal variation of the oxygen
  abundances that increases the dispersion when not taken into
  account. Finally, we study if there is a dependence with
  galactocentric distance. We found that the mixing scale-length is
  slightly lower in the inner regions $r_{mix}(r/re<0.9)=$0.28$r_e$ than
  in the outer ones $r_{mix}(r/re>0.9)=$0.67$r_e$.

 To know how sensitive  this
  dispersion is to the errors and uncertainties in the derived parameter,
  we performed a simple Monte-Carlo simulation, allowing each of the
  parameters (abundances, galactocentric distances, effective radius
  and inclination) to vary within the estimated errors. The standard 
  deviation between the different estimated radial mixing scales is
  $\sim$0.15$r_e$.

 }

This parameter puts a strong
constraint on the metal mixing scale-length, independently of the
mechanism required to produce the mixing.   Obviously this is an upper limit to the
mixing scale-length, in particular if the abundance gradient depends
on the equivalent width of H$\alpha$.

\begin{figure}
\centering
\includegraphics[width=7.2cm,angle=270]{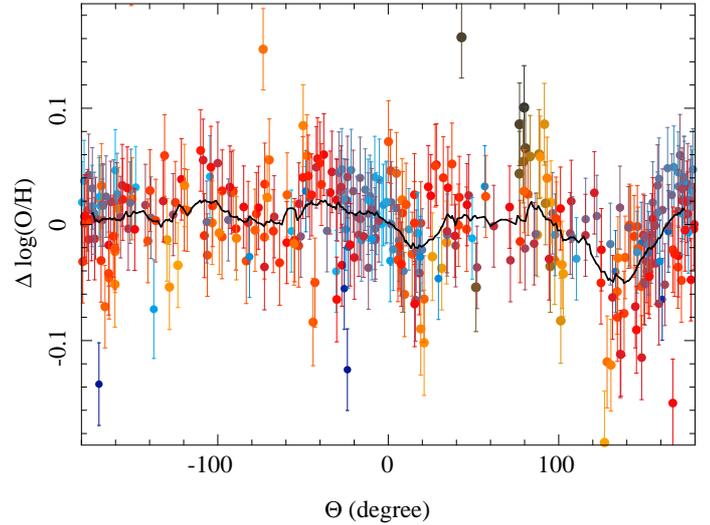}
\caption{\label{fig:AZ}  Azimuthal distribution of the residual of the oxygen abundance for 
the individual \HII regions once subtracted the average radial gradient. The colors and sizes
indicate the galactocentric distance, with blue solid circles corresponding to \HII
regions more nearer to the center of the galaxy, and red/orange ones corresponding to those 
 farther away. The solid line corresponds to the average value at each azimuthal angle for \HII
regions within 25\degree of the considered angle.}
\end{figure}

{ 
\subsection{Azimuthal variations of the oxygen abundance}\label{OH_AZ}

In a pure inside-out scenario where the metal enrichment is dominated
by local processes (the metal pollution by stars that dies at a
certain location), the abundance gradient should not present any
azimuthal variation.  Different mechanisms proposed for the radial
mixing predict different characteristic patterns in the azimuthal
distribution of the oxygen abundance. The sample of \HII regions
provided by our MUSE data is large enough to explore if there is an
azimuthal variation in the distribution of oxygen abundances.

Figure \ref{fig:AZ} shows the azimuthal distribution of the oxygen
abundances for the \HII regions of our catalog, once removed the
common radial gradient, for those regions within a ring of
0.3$<r/r_e<$2.1 (i.e., the linear regime of the abundance gradient).
There is a clear pattern, more clear when we derive
the azimuthal average within a box of 25\degree around each value.
Indeed, when removing this average pattern the dispersion around the
mean value is reduced by a 40\%. The amplitud of the pattern does not
follow a clear periodic sequence (like a sinusoidal structure), and
there is no clear general dependence with the galactocentric distance.

However, in the strongest feature of this pattern, the wiggle between
$\theta\sim$90\degree and $\theta\sim$160\degree with an amplitude of $\sim$0.05 dex, there seems to be a
trend with distance, with the abundance decreasing at intermediate
distances and increasing for both the inner and outer regions. This pattern
corresponds to the spiral arm the south-east of the center of the galaxy.
If real, this could be a hint of radial mixing.

The radial mixing scale defined in the previous section is reduced to
0.39$r/re$, or 4.1kpc, when the average azimuthal variation of the
oxygen abundance is subtracted prior to derive the dispersion around
the radial abundance gradient.

}

\section{Discussion and Conclusions}\label{discuss}

In this study we analyse one of the first observations using MUSE
on a spiral galaxy, NGC 6754. We detect and extract the spectroscopic
information of a sample of 396 \HII regions, an order of
magnitude larger than the average number observed in previous state-of-the-art
IFU survey datasets \citep[e.g.][]{sanchez14}.  This
illustrates the capabilities of this unique instrument due to the
combination of its large FoV and unprecedent spatial
sampling/resolution.

The abundance distribution derived has a negative gradient, with a
slope consistent with the characteristic value reported by previous
studies \citep{sanchez14}, { in the linear regime between 0.3 and 1.7
  $r_e$. The abundance decreases in the inner regions, and there is a hint of
a flatenning in the outer parts. Both features have been already
  observed in previous studies in individual galaxies
  \citep[e.g.,][]{bresolin09,yoac10,rosales11,mari11,bresolin12}. The central drop is in
  many cases associated with a circumnuclear star-formation process
  \citep{sanchez14}, and it could be related with the accumulation of
  gas due to non-circular motions exerted near the inner Lindblad
  resonance radius \citep[e.g.,][]{cepa90}. 

The nature of the flatenning is still not clear. A detailed discussion
on the different scenarios proposed was presented by \cite{sanchez14}.
In summary, under the usually observed star-formation rates, the time
required to enrich the ISM up to the observed abundances in these outer
regions is of the order of the age of the Universe \citep{bresolin12},
and therefore it is unlikely that {\em in situ} star-formation could
have enriched the interstellar medium to the values observed. 
Among the main mechanisms proposed to explain the flatenning we
highlight the following ones: (i) angular momemtum transport that
produces a radial mixing
\citep[e.g.,][]{1985ApJ...290..154L,1992A&A...262..455G,2000A&A...355..929P,2009MNRAS.396..203S,2011A&A...531A..72S};
(ii) resonance scattering with transient spiral density waves
\citep{2002MNRAS.336..785S}; (iii) the overlap of spiral and bar
resonances \citep{2011A&A...527A.147M}; (iv) stellar radial migration
\citep[e.g.,][]{2008ApJ...675L..65R,2008ApJ...684L..79R}; and (v)
minor mergers and captures of satellite galaxies
\citep{2009MNRAS.397.1599Q,2012MNRAS.420..913B}.

We explore the possible dependence of the slope of the abundance gradient
with different properties of the ionized gas. 
Significant differences are not expected if the metal enrichment is dominated
by the inside-out growth of the galaxy. However, local processes, like outflows or metal raining induced
by enhanced star-formation associated with the spiral arms may
modify the chemical distribution locally. Under this assumption it would be
expected that denser \HII regions in spiral arms, with greater \EWHa, 
and ionization strengths  present a different distribution
of oxygen abundances that those \HII regions located in the inter-arm regions. 
Our results indicate that local processes do not seem to be relevant enough
to modify the galactocentric abundance gradient in this particular galaxy.

}

We { define a parameter to estimate the amount of the redistribution of metals within 
the galaxy that we call the mixing scale-length, $r_{mix}$. This parameter is defined as the dispersion around the abundance gradient along the galactocentric distance, and can be derived as the ratio between the dispersion in the abundance and the slope of the correlation.} We estimate the typical $r_{mix} =$0.43$\pm$0.15 $r_e$, i.e., $\sim$4.6 kpc at the redshift of the galaxy. { To our knowledge, this is the first time that this parameter is defined in this way. However, it is possible to compare with previous results if both the dispersion in abundance and the slope of the radial gradient are provided. The most recent exploration of the abundance gradient over a large sample of galaxies was published by \cite{sanchez12b,sanchez14}. They found that the common abundance gradient has a slope of $\alpha=-$0.1 dex/$r_e$, with a dispersion of $\sim$0.6 dex, for which an average mixing scale-length of $r_{mix}\sim$0.6 $r_e$ is derived. This value is slightly larger than the one we find for NGC 6754. However, we should note here that the estimation derived from \cite{sanchez12b,sanchez14} results is purely statistical, based on the average abundance gradient derived once the individual gradients for each galaxy are considered all together and normalized to the abundance at the effective radius. Measurements on individual galaxies have not been provided.} 

{ 
Bars have been proposed as an effective mechanism for radial migration
\citep[e.g.,][]{atha92,sellwood02}. Hydrodynamical simulations have shown that
bars induce angular momentum transfer via gravitational torques, that
result in radial flows and mixing of both stars and gas
\citep[e.g.,][]{atha92}.
 These radial motions can produce a mixing and
homogenization of the gas, that leads to a flattening of any
abundance gradient \citep[e.g.,][]{frie98}. Resonances  between
the bar and the spiral pattern speeds can shift the orbits of stars,
mostly towards the outer regions \citep{minc10}, a mechanism that
also affects the gas. Another process that produces a similar effect is
the coupling between the pattern speed of the spiral arms and the bar,
that induces angular momentum transfer at the corotation radius
\citep[e.g.,][]{sellwood02}. In a recent study \cite{dimat13} analysed the
signatures of radial migration in barred galaxies on the basis of simulations. They found that the
slope of the abundance gradient does not change significantly up to $\sim$1.5-2 $r_e$ (when
the scale-length of their simulated disks are transformed to an effective radius), 
but a flatenning is predicted beyond these galactocentric distances. This pattern
is very similar to the one observed in our galaxy. However, as we discussed before, the flatenning
in the outer region seems to be present in spiral galaxies irrespective of the presence or absence of bars
\citep{sanchez14}.

\cite{dimat13} defined a parameter to quantify the amount of spatial
redistribution of stars in a disk as the ratio between the maximum
absolute variation of the metallicity with respect to the radial
gradient compared to the slope of this gradient
($\delta_{[Fe/H]}/\Delta_{[Fe/H]}$). With units of distance, this
parameter is equivalent to our mixing scale-length ($r_{mix}$). They
found that this parameter evolves with time and presents a weak radial
dependence. At the peak of the radial migration it ranges between
1-1.5 kpc (0.17-0.26 $r_e$) for a galactocentric distance between 3
and 12 kpc (0.5-2 $r_e$). As time evolves it decreases, being slightly
larger in the outer regions. Our derived $r_{mix}$ is larger, but certainly of the same order, 
than that predicted by \cite{dimat13}. It also
presents a weak radial dependence, that may indicate that the peak of the
migration has already past. This is a good agreement
considering that we are not comparing with {\it ad hoc} simulations
specifically done  to reproduce our galaxy.

Another prediction by simulations is that the radial mixing should not
be homogeneous. These inhomogeneities are related to the way radial
migration occurs in galaxies, following the arms pattern
\citep[e.g.,][]{min12}: metal-rich stars which move to the outer disk
are mostly from the region outside corotation \citep{brun11}, and
migrate through spiral patterns to the outer parts of the disk. In
other words, migration is not axisymmetric, but associated to the
distribution of arms and bars. In Section \ref{OH_AZ} we explored the
possible azimuthal variations of the oxygen abundance, once subtracted
the radial dependence, and we found evidence of an asymmetrical
distribution. The strongest feature is associated with the spiral arms
in the south-east of the galaxy.  The amplitude seems to be smaller
than that predicted by \cite{dimat13}, for the epoch of the strongest
migration: $\sim$0.2 dex at $\sim$7 kpc (1 $r_e$), at t=1.1 Gyr in
their simulations (Fig. 8, top panel, in that article). However, 
this effect is expected to become weaker with time (Figure 8, bottom panel
of that article), and our previous result indicates that this galaxy has
already passed the peak of the strongest migration.

The proposed scenario assumes that the deviation of the abundances
with respect to the radial gradient, due to the radial migration associated
with arms and bars, should be stronger in barred galaxies. This is a consequence of the
stronger radial movements expected to be induced by these morphological features. In this context
it is interesting to note that recent results indicate that the stellar and gas
kinematics of barred and un-barred galaxies seem to be very similar, without stronger
distorsions induced by the presence of the bar, at least at large scales \cite{jkbb14}.

}

{ Despite the fact of this significant advance in our understanding
  of the possible effects of radial mixing, it is important to
  highlight that all these results were derived for a single
  galaxy}. Larger samples of galaxies are needed to explore { whether
  the estimated mixing scale-lengths and azimuthal variations depend
  on other properties of the galaxies}, such as the presence or
absence of bars, { the strength of the bars}, the interaction
stage, the morphological type, and the stellar mass or luminosity.
{ In particular, if the picture outlined by \cite{dimat13} is valid,
  we should find different strengths in both $r_{mix}$ and the
  intensity of the azimuthal variations depending on the timing of the
  evolution of the bars and the coupling or not with the spiral arms.
  Another important caveat is that the results from the current
  simulations were focused on the effects of the migration on old
  stars, and in their metallicities.  As clearly illustrated in 
  recent results by \cite{rosa14b} the gas-phase abundance is better
  correlated with the metallicity of young stars ($t<$2 Gyr), with old
  stars being in general more metal poor. It is still unclear how these
differences may affect the interpretation of our results on the basis of the
simulations. However, there is a lack of similar simulations on the effects
of radial migration for the gas-phase abundance.}

 The current { results} illustrate the capabilities of MUSE to
 accomplish this kind of studies in a very efficient way, { and
   demonstrate that it is possible to derive reliable dispersions
   around the mean abundance gradient. In future articles we will
   apply the methodology outlined here to a sample of galaxies with
   similar characteristics of cosmological distances and projected
   sizes, observed with this instrument, in order to explore the
   dependence of the  results on galaxy type, as outlined before. }

\begin{acknowledgements}

SFS thanks the director of CEFCA, M. Moles, for his sincere support.

We thank the referee for his/her comments that have improved this manuscript.

Based on observations made with ESO Telescopes at the Paranal Observatory under programme ID 60.A-9329.
Support for LG and HK is provided by the Ministry of Economy, Development, and Tourism's Millennium Science Initiative through grant IC12009, awarded to The Millennium Institute of Astrophysics, MAS. LG and HK acknowledge support by CONICYT through FONDECYT grants 3140566 and 3140563, respectively. SFS acknowledges the Mexican National Council for Science and Technology (CONACYT) for financial support under the program {\it Proyectos de Ciencia Basica}. EP acknowledges funding from the Spanish MINECO grant AYA2010-15081. RA Marino was also funded by the Spanish programme of International Campus of Excellence Moncloa (CEI).

\end{acknowledgements}

\bibliography{CALIFAI}
\bibliographystyle{aa}

\end{document}